\documentclass[twocolumn]{aa}
\usepackage{graphicx,amssymb}
\newcommand \msun {\mbox{$\mathcal{M}_{\odot}$}}
\newcommand \kms {\mbox{km~s$^{-1}$}}
\newcommand \degree {\mbox{$^\circ$}}
\newcommand \lsun {$L_{\odot}$\ }

\begin{document}
   \title{NUclei of GAlaxies}

   \subtitle{V.\ Radio emission in 7 NUGA sources\thanks{EVN: The
   European VLBI Network is a joint facility of European, Chinese,
   South African and other radio astronomy institutes funded by their
   national research councils. MERLIN is a national facility operated
   by the University of Manchester on behalf of PPARC. VLBI including the
   VLBA: The National Radio Astronomy Observatory is a facility of the
   National Science Foundation operated under cooperative agreement by
   Associated Universities, Inc.}}

   \author{M.\ Krips\inst{1,2},
           A.\ Eckart\inst{1},
	   T.P.\ Krichbaum\inst{3},
           J.-U.\ Pott\inst{1,4},
           S.\ Leon\inst{5,6},
           R.\ Neri\inst{7},
           S.\ Garc\'{\i}a-Burillo\inst{8},
           F.\ Combes\inst{9},
           F.\ Boone\inst{9},
           A.J.\ Baker\inst{10},
           L.J.\ Tacconi\inst{11},
           E.\ Schinnerer\inst{12}, 
	   \and
           L.K. Hunt\inst{13}
%          P. Englmaier\inst{12}
          }

   \offprints{M.Krips\\
              e-mail: mkrips@cfa.harvard.edu}

  \institute{Universit\"at zu K\"oln, I.Physikalisches Institut,
              Z\"ulpicher Str. 77, 50937 K\"oln, Germany;\\
              \email{krips@ph1.uni-koeln.de,eckart@ph1.uni-koeln.de}
            \and
            Harvard-Smithsonian Center for Astrophysics, SMA project, 
	    645 North A`ohoku Place, Hilo, HI 96720;\\
              \email{mkrips@cfa.harvard.edu}
            \and
              Max-Planck Institut f\"ur Radioastronomie,
	      Auf dem H\"ugel 69, 53121 Bonn, Germany;\\
              \email{tkrichbaum@mpifr-bonn.mpg.de}
            \and
              ESO, Karl-Schwarzschild-Str. 2, 85748 Garching,
              Germany;
              \email{jpott@eso.org}
            \and
              Instituto de Astrof\'{\i}sica de Andaluc\'{\i}a (CSIC),
              C/ Camino Bajo de Hu\'etor,24, Apartado 3004, 18080 Granada,
              Spain;
             \and
              IRAM, Avenida Divina Pastora, 7, N\'ucleo Central,
              18012 Granada, Spain;
              \email{leon@iram.es}
             \and
              Institut de Radio-Astronomie Millim\'etrique (IRAM),
              300, rue de la Piscine, 38406 St.Martin-d'H\`eres,
              France;\\
              \email{neri@iram.fr}
            \and
              Observatorio Astron\'omico Nacional (OAN)-Observatorio
              de Madrid, Alfonso XII, 3, 28014 Madrid, Spain;\\
              \email{s.gburillo@oan.es}
            \and
             Observatoire de Paris, LERMA, 61 Av. de l'Observatoire,
              75014 Paris, France;\\
              \email{francoise.combes@obspm.fr,fboone@mpifr-bonn.mpg.de}
            \and
              Department of Physics and Astronomy, Rutgers, the State 
              University of New Jersey, 136 Frelinghuysen Road, Piscataway,
              NJ 08854-8019, United States;\\
              \email{ajbaker@physics.rutgers.edu}
            \and
              Max-Planck-Institut f\"ur extraterrestrische Physik,
              Postfach 1312, 85741 Garching, Germany;\\
              \email{linda@mpe.mpg.de}
	    \and  
	      Max-Planck-Institut f\"ur Astronomie,
	      K\"onigstuhl 17, 69117 Heidelberg, Germany;
              \email{schinner@mpia.de}
            \and
              INAF-Istituto di Radioastronomia/Sezione Firenze, Largo
              E.\ Fermi, 5, 50125 Firenze, Italy;\\
              \email{hunt@arcetri.astro.it}
%            \and
%              Astronomisches Institut, Universit\"at Basel,
%              Venusstr. 7, 4102 Binningen, Switzerland;\\
%              \email{ppe@astro.unibas.ch}
%             \thanks{wem auch immer}
             }

   \date{Received ; accepted}

   \abstract{We present high angular resolution radio snap-shot
  observations of seven nearby low-luminosity active galaxies (LLAGN)
  from the NUclei of GAlaxies (NUGA) survey.  The observations were
  conducted with MERLIN and EVN/VLBI at 18~cm and 6~cm. At all
  observed angular resolutions and frequencies, we find indications
  for extended emission in about $\sim$ 40\% of the sources,
  consistent with the decrease of flux with increasing angular
  resolution. The extended components resemble jet emission in a
  majority of cases, consistent with the optically thin synchrotron
  emission implied by their steep spectra. We consider the compact 6cm
  EVN/VLBI radio emission of our sources in the context of the
  ``fundamental plane'' that previous LLAGN studies identified within
  the three-dimensional parameter space of radio luminosity, X-ray
  luminosity, and black hole mass. We demonstrate, using NGC~7217 and
  NGC~1068 as particular examples, that high-resolution, multi-epoch
  radio observations offer useful information about the origin of
  offsets from the fundamental plane.

   }

\authorrunning{M.Krips}

\maketitle
%
%________________________________________________________________

\begin{table*}[!t]
\centering
\begin{tabular}[c]{lcccccc}
    \hline
    \hline
    Name    &  RA       & DEC     & v$_{\rm hel}$ & $i$$^a$ & Host & Nuclear \\
            &  (J2000)  & (J2000) & [\kms]       & (\degree)& type & type    \\
    \hline
    NGC1961 &  05:42:04.6 &  69:22:43.0 & 3934 & 49 & SAB(rs)c    & L2    \\
    NGC2782 &  09:14:05.1 &  40:06:49.0 & 2562 & 42 & SAB(rs) pec & H     \\ 
    NGC3147 &  10:16:53.6 &  73:24:03.0 & 2820 & 26 & SA(rs)bc    & S2    \\ 
    NGC3718 &  11:32:35.0 &  53:04:04.0 & 994  & 60 & pec      & L1.9  \\   
    NGC4579 &  12:37:43.4 &  11:49:05.0 & 1519 & 37 & SAB(rs)b   & S1.9/L1.9\\ 
    NGC5953 &  15:34:32.4 &  15:11:38.0 & 1965 & 36 & SAa pec     & S2    \\ 
    NGC7217 &  22:07:52.4 &  31:21:34.0 & 952  & 35 & (R)SAB(rs)a & L2    \\  
    \hline
\end{tabular}
\caption{Basic properties of the NUGA sources taken from NED. $^a$
  $i\equiv$ inclination of the galaxy. H\,=\,HII-region, S\,=\,
  Seyfert and L\,=\,LINER.}
\label{nugaprop}
\end{table*}

\section{Introduction}

Nuclear activity in galaxies is detected at many different
levels, supporting the idea that it reflects different phases in
galaxy evolution, i.e., that every galaxy might exhibit nuclear
activity at some point in its life. Nuclear activity is thought to be
due to accretion onto supermassive black holes (SMBHs), which are now
accepted to reside in virtually all galaxies, at least those with
bulges (Kormendy and Richstone 1995; Magorrian et al.\ 1998). The
tight correlation of SMBH mass with the stellar velocity dispersion in
the bulge (Magorrian et al.\ 1998; Richstone et al.\ 1998; Gebhardt et
al.\ 2000; Ferrarese \& Merritt 2000; Tremaine et al.\ 2002; Shields
et al.\ 2003; Marconi \& Hunt 2003) along with the similar redshift
peaks ($z \approx 2$) of the quasar and starburst populations,
suggests a connected, or at least similar evolution between starburst
and SMBH activity. As implied by the population peak, nuclear activity
seems to increase with redshift, leaving predominantly Low-Luminosity
AGN (LLAGN) and inactive galaxies in the nearby universe. However,
many similarities are found between LLAGN and High-Luminosity AGN
(HLAGN) over the entire electromagnetic spectrum, further indicating
that different activity phases might simply correspond to different
evolutionary stages.

Recent studies strongly suggest that HLAGN, LLAGN, and even X-ray
binaries occupy a so-called ``fundamental plane'' within the
three-dimensional parameter space of central black hole mass vs. radio
luminosity vs. X-ray luminosity (Merloni, Heinz \& DiMatteo 2003;
Falcke, K\"ording \& Markoff 2004).  This strongly indicates common
underlying processes.  The {\it exact} origin of this correlation is
still the subject of controversy, although it is now widely accepted
that different accretion rates and/or different accretion efficiencies
certainly play a significant role in setting the various activity
levels. In this picture, LLAGN may be driven by low accretion rates
into a mode in which they only release a small fraction of their
energy through radiation from a radiatively inefficient accretion flow
(RIAF); the remainder is instead converted into thermal, turbulent, or
mechanical energy. HLAGN, however, efficiently cool through radiation,
thus producing a high activity level. A variety of RIAF models have
been invoked over the past decade, including Bondi-Hoyle accretion
(Melia \& Falcke 2001), Advection Dominated Accretion Flows (ADAFs;
e.g., Narayan et al.\ 1995), Convection Dominated Accretion Flows
(CDAFs; Ball et al.\ 2001), ADiabatic Inflow Outflow Solutions (ADIOS;
Blandford \& Begelman 1999) and quasi-mono-energetic electron
distributions (Beckert \& Duschl 1997). However, ADAFs and Bondi-Hoyle
accretion are found to fail for Sgr~A$^\star$ in the center of our own
galaxy, since they entail gas densities which are significantly higher
than observed; they are thus already regarded as rather
out-of-date\footnote{For a more detailed description of the different
RIAFs we refer to Quataert et al.\ (2003), Eckart et al.\ (2004) and
the references given above}.

As an alternative to RIAFs, a disk-jet coupling can also produce the
observed fundamental plane (Heinz \& Sunyaev 2003); here, the base of
the jet rather than a RIAF/disk produces the compact (optically thick
synchrotron) emission in LLAGN, while the accretion disk dominates the
emission in HLAGN (e.g., Falcke \& Markoff 2000). Previous studies of
LLAGN (e.g., Ulvestad \& Wilson 1984; Pedlar et al.\ 1993; Falcke et
al. 2000; Nagar et al.\ 2002, 2005; Anderson, Ulvestad \& Ho 2004;
Anderson \& Ulvestad 2005) indicate radio jets and/or compact radio
emission similar to but less powerful than in HLAGN. Jets are
intimately connected to accretion onto compact objects (e.g., Pringle
1993; Blandford 1993) and are thus regarded as a clear signature of
accretion-powered AGN. However, it is impossible at the moment to
distinguish between emission powered by a jet and a RIAF, so various
combinations of the two must generally be considered (e.g., Yuan et
al.\ 2002). In each case, high angular resolution observations of the
radio emission offer substantial insight into the nature of the
underlying process.

The seven sources studied in this paper belong to the IRAM NU(clei of)
GA(laxies) survey (Garc\'{\i}a-Burillo et al.\ 2003). The NUGA project
aims at studying the molecular gas distributions and kinematics with
the highest currently available angular resolution and sensitivity in
a sample of 12 nearby Seyfert and LINER galaxies and transition
objects (Garc\'{\i}a-Burillo et al.\ 2003; Combes et al.\ 2004; Krips
et al.\ 2005; Garc\'{\i}a-Burillo et al.\ 2005). The study of radio
emission is an ideal complement to the molecular gas analysis in
investigating the activity in these sources. We thus selected those
sources in our sample with existing and clear VLA radio detections
(Becker, White \& Helfand 1995; Condon et al.\ 1996; Ho \& Ulvestad
2001) for a higher angular resolution study with MERLIN and VLBI.

The paper is organized in the following way. Section~\ref{obs} gives
an overview of the observations. Section~\ref{obsres} describes the
obtained results. The nature of the detected radio emission is
discussed in Section~\ref{sec-radio-sed}. A comparison of the radio
luminosities with respect to the fundamental plane is presented in
Section~\ref{corr}. Finally, the paper ends with a summary and
conclusions in Section~\ref{sum}.

%__________________________________________________________________

\section{Observations}
\label{obs}
Radio snapshot\footnote{snapshot corresponds here to $\sim$3-5 hours
on-source integration time.} observations of seven NUGA sources
(Table~\ref{nugaprop}) have been carried out at 18~cm and 6~cm using
MERLIN and EVN/VLBI between 2001 and 2004. Two sources from this
sample (NGC~3718 and NGC~3147) have been used for deeper integration
(6-9~hours per source). The history of the observations is listed in
Table~\ref{nuga-obsepoch}. We selected these seven sources from the
NUGA survey because they have all been clearly detected on arcsecond
scales before (VLA observations: Becker, White \& Helfand 1995; Condon
et al.\ 1996; Ho \& Ulvestad 2001) but lacked consistent mapping at
higher angular resolution.  With both MERLIN and EVN/VLBI, we took
advantage of the phase-reference technique (Beasley \& Conway 1995) by
observing alternately the target and a phase calibrator located within
a few degrees. The EVN/VLBI observations were conducted with the Mark
IV system at a rate of 256~Mb$\,{\rm s}^{-1}$ using 8 IFs, each of
8~MHz bandwidth, and 2-bit sampling (mode-256-8-2). The sources were
observed in LCP. At MERLIN, polarization measurements were also
carried out but without any significant detection. The visibility
amplitudes were calibrated assuming Kelvin to Jansky conversion
factors, gain-elevation dependencies, and system temperatures measured
during the observations.  This and careful amplitude self-calibration
using the available closure amplitudes provided a typical calibration
accuracy of $5-10$\,\%. Approximate beam sizes range from 0.05 to
0.5$''$ for MERLIN and 0.001 to 0.04$''$ for EVN. RMS image noises
range from $\sim$1~$\mu$Jy to $\sim$1~mJy\footnote{The lower rms noise
corresponds to the two deep integrations, not the snap-shot
observations}.  All data were reduced with the AIPS package and mapped
either in AIPS itself or using DIFMAP (van Moorsel et al.\ 1996;
Shepherd 1997).

Additionally, we have information on the continuum emission at 3~mm
and 1~mm for all seven sources, which was gained during the CO(1--0)
and CO(2--1) observations of the NUGA sample at the IRAM Plateau de
Bure Interferometer (PdBI). For a more detailed description of the
PdBI observations, we refer the reader to previous NUGA publications
(e.g., Garc\'{\i}a-Burillo et al.\ 2003; Combes et al.\ 2004; Krips et
al.\ 2005; Garc\'{\i}a-Burillo et al.\ 2005).

\begin{table*}
\centering
\begin{tabular}[c]{ccccl}
    \hline
    \hline
    Instrument & Obs. Frequency  & Month - Year & VLBI stations$^d$ 
 & Sources\\
    \hline
    MERLIN      & 1.66GHz  & 2001/2002    & - & all except NGC~4579$^a$ \\
    EVN         & 1.63GHz  & February - 2002  & Eb,Jb,Wb,Mc,Nt,
    & NGC~3147, NGC~3718 \\
    & & & On85,Sh,Ur,Tr  &\\
    EVN         & 1.63GHz  & November - (2002) 2003$^c$ 
    & Eb,Jb,Wb,Mc,Nt, &  NGC~1961, NGC~2782, NGC~4579, \\
    & & & On85,Sh,Ur,Tr  & NGC~5953, NGC~7217    \\
    MERLIN      & 4.99GHz  & 2002/2003    &-  & all                     \\
    EVN         & 4.99GHz  & June - 2003  &  Eb,Jb,Wb,Mc,Nt,
    & NGC~3147, NGC~3718      \\ 
    & & & On85,Sh,Ur,Tr  &\\
    global VLBI & 4.99GHz$^b$  & May - 2004   &  Eb,Jb,Wb,Mc,Nt,
    &NGC~1961, NGC~2782, NGC~3718,\\
    & & & On85,Sh,Ur,Tr,VLBA & NGC~5953, NGC~7217  \\
    \hline
    \hline
\end{tabular}
\caption{Radio observations for seven NUGA sources. $^a$ NGC4579 was
excluded since MERLIN observations were already carried out; the data
are taken from the MERLIN archive. $^b$ NGC~3147 and NGC~4579 were not
observed because VLBA data were already available from Ulvestad \& Ho
(2001). $^c$ The first observations in 2002 were not successful due to
technical problems with some of the antennas and were thus repeated in
2003; the results of both observations are consistent with each other
but only the 2003 data sets are used for further discussion due to the
higher quality of the data. $^d$ Eb$\equiv$Effelsberg,
Jb$\equiv$Jodrell Bank,Wb$\equiv$Westerbork,
Mc$\equiv$Medicina,Nt$\equiv$Noto,On85$\equiv$Onsala 25m,
Sh$\equiv$Shanghai, Ur$\equiv$Urumqi,Tr$\equiv$Torun (see EVN homepage
for more information on the respective telescopes:
http://www.evlbi.org/).}
\label{nuga-obsepoch}
\end{table*}

\section{Results}
\label{obsres}
We fitted Gaussian profiles in the image plane to all of our radio
maps to get estimates of component sizes (by using the fitted FWHM of
the continuum emission and half the beamsize as a cutoff lower limit)
and to derive the respective fluxes (Table~\ref{radio-pos} \&
\ref{radio-fluxes}) and positions (Table~\ref{radio-pos}). We first
start with some general results for the entire sample in
Section~\ref{genres} and then discuss each galaxy individually in
Section~\ref{indres}.

\subsection{General results}
\label{genres}
Tables~\ref{radio-pos} and \ref{radio-fluxes} list the derived
positions, fluxes and sizes of the respective sources. The position of
each component is consistent across all data sets.
Figs.~\ref{1st-3-radio} to \ref{NGC7217-radio} show the continuum maps
of the seven sources for each frequency and instrument. We find
extended emission on all angular scales. This is confirmed by
comparing the peak intensities and integrated flux densities with each
other (a difference indicating resolved emission). Extended emission
is also implied by the decrease of flux density as we move from VLA to
MERLIN to EVN resolution at a fixed frequency (e.g., in NGC\,1961 and
NGC\,7217; see Table~\ref{radio-lum} for the VLA luminosities). In at
least three objects (NGC~2782, NGC~5953, and NGC~7217), we see clear
signs of a jet on MERLIN scales, and in another two (NGC~1961 and
NGC~3718) we find indications of jet-like morphologies. Interestingly,
while for the stronger sources we find flat to inverted spectra, the
weaker sources have steep spectra (Fig.~\ref{specind} and
Table~\ref{radio-specind}).

\subsection{Results on individual galaxies}
\label{indres}
In the next subsections we will present separately the individual
results for each galaxy. 

We determine for all galaxies the nuclear bolometric luminosities via
the X-ray flux unless otherwise stated. Ulvestad \& Ho (2001) propose
an empirical relation to estimate the bolometric luminosity of the
nucleus in active galaxies.  Based on a sample of 10 objects, they
find $L_{\rm bol}=6.7\times L_{\rm x}$(2-10keV). Assuming this
equation and taking the X-ray luminosity (Table~\ref{radio-lum})
published by Roberts \& Warwick (2000), Ulvestad \& Ho (2001),
Fabbiano et al.\ (1992), or Terashima et al.\ (2002), we can crudely
estimate the nuclear bolometric luminosity for each galaxy. In some
cases, this estimate should be taken as an upper limit: particularly
for Roberts \& Warwick (2000), a significant contribution of
circumnuclear star formation to the X-ray flux cannot be entirely
excluded, due to the low ($\sim 10^{\prime\prime}$) angular resolution
of the observations. Furthermore, we have hard (2-10\,keV) X-ray
luminosities for only three sources, requiring that we extrapolate the
hard X-ray flux from the soft 0.2-4\,keV (or 0.2-2\,keV) band for the
remaining sources as a crude estimate. These values are only regarded
as rough estimates of the nuclear hard X-ray luminosities, given
significant uncertainties in the contribution of emission from the
host galaxy, the spectral shape in this band (sometimes a break occurs
at $\sim$1~keV), and the amount of absorption. The last-mentioned
uncertainty might be the most important one, as all of the sources
studied in this paper are classified as type 2 Seyfert or LINER,
implying an obscuration of the central engine in the standard AGN
unification paradigm. However, previous X-ray studies of type 2
Seyfert and LINER galaxies, including NGC~3147, NGC~4579, and NGC~7217
(e.g., Terashima et al.\ 2002), indicate that even these type 2 LLAGN
typically show spectral indices of $\Gamma\simeq$1.6-2.0 (with
P(E)=A$\cdot$E$^{-\Gamma}$), suggesting that our approximation
($\Gamma\geq$1) might give a reasonable and reliable upper limit.

We estimate black hole masses for all galaxies using the $M_{\rm
bh}$-$\sigma_{\rm s}$ correlation, with $\sigma_{\rm s}$ being the
stellar velocity dispersion of the bulge component (e.g., Gebhardt et
al.\ 2000; Tremaine et al.\ 2002). The velocity dispersion has been
taken from McElroy et al.\ (1995) for almost all galaxies except for
NGC3718, for which we estimated it from {\small FWHM(N\,[II])} taken
from Ho, Filippenko \& Sargent (1997; see also caption of
Table~\ref{radio-lum}). As a caveat, one has to keep in mind that the
$M_{\rm bh}$-$\sigma_{\rm s}$ correlation has been mostly established
for elliptical galaxies, while our sample is based on late type spiral
galaxies. However, recent results from reverberation mapping indicate
that the relation still holds for spiral galaxies containing a Seyfert
nucleus (e.g., Wandel et al. 1999; Kaspi et al. 2000; Nelson et al.\
2004; Onken et al.\ 2004; Peterson et al.\ 2005). A different,
recently significantly improved approach to determine the black hole
mass is the bulge-to-black-hole mass correlation (e.g., Marconi \&
Hunt 2003, H\"aring \& Rix 2004). An estimate of the bulge masses for
the NUGA sample is in preparation (Hunt et al.) so that we refer the
reader to this paper, which will also include a comparison of the
various black hole mass estimates.

\begin{table*}
\centering
\begin{tabular}[c]{lllllllll}
  \hline
  \hline
  Name     & RA [h:m:s] & $\Delta\alpha$[s] 
           & Dec [\degree:$'$:$''$] & $\Delta\delta$[$''$] \\ 
  \hline
  NGC1961-core & 05:42:04.6477 & $\pm$0.0003  & 69:22:42.375  & $\pm$0.004 \\ 
% NGC1961-jet  & 05:42:04.67   & $\pm$0.01    & 69:22:42.4    & $\pm$0.2    \\ 
  NGC2782-core & 09:14:05.1124 & $\pm$0.0003  & 40:06:49.316  & $\pm$0.004 \\
  NGC2782-jet  & 09:14:05.105  & $\pm$0.001   & 40:06:49.27   & $\pm$0.01  \\
  NGC3147-core & 10:16:53.6500 & $\pm$0.0003  & 73:24:02.680  & $\pm$0.004 \\
  NGC3718-core & 11:32:34.8530 & $\pm$0.0003  & 53:04:04.518  & $\pm$0.004 \\ 
  NGC3718-jet  & 11:32:34.854  & $\pm$0.001   & 53:04:04.523  & $\pm$0.004 \\ 
  NGC4579-core & 12:37:43.522  & $\pm$0.001   & 11:49:05.498  & $\pm$0.004 \\
  NGC5953-core & 15:34:32.383  & $\pm$0.002   & 15:11:37.59   & $\pm$0.02 \\
  NGC5953-jet  & 15:34:32.390  & $\pm$0.003   & 15:11:38.03   & $\pm$0.02 \\
  NGC7217-core & 22:07:52.3933 & $\pm$0.0003  & 31:21:33.646  & $\pm$0.004 \\
  \hline
\end{tabular}
\caption{Radio positions of the seven NUGA sources, mainly derived
  from the EVN 6cm (core) or MERLIN 18cm/6cm (jet) observations by
  fitting Gaussian profiles to the emission (the positions are
  coincident within the uncertainties of the different
  observations). The positional uncertainties are estimated from the
  Gaussian fits and also include the most recent (i.e., determined in
  the past 4 years) astromectric errors of the used
  phase-calibrators. }
\label{radio-pos}
\end{table*}

\begin{table*}
\centering
\begin{tabular}[c]{lcccccc}
    \hline
    \hline
             &\multicolumn {6}{c}{\hrulefill\hskip 3mm
    \raisebox{-1.0mm}{MERLIN}\hskip 3mm \hrulefill} \\
     & \multicolumn{3}{c}{18cm}
     & \multicolumn{3}{c}{6cm}\\
    Name          & Peak & Integrated & Deconvolved 
                  & Peak & Integrated & Deconvolved \\
                  & flux & flux density & Size  
                  & flux & flux density & Size \\
                  & [mJy/beam] & [mJy] & $''\times''@\degree$  
                  & [mJy/beam] & [mJy] & $''\times''@\degree$  \\
    \hline
    NGC1961-core & 2.2$\pm$0.2  & 3.3$\pm$0.4  & 0.1$\times$0.09@143
                  & 1.2$\pm$0.3  & 1.8$\pm$0.6  & 0.1$\times$0.1 \\
%   NGC~1961-core & 2.3$\pm$0.2 & 2.4$\pm$0.3   & 0.06$\times$0.08@150 
%                 & 1.2$\pm$0.3 & 1.8$\pm$0.6   & 0.3$\times$0.3@170 \\
%   NGC~1961-jet  & 0.4$\pm$0.2 & 2.3$\pm$1.0   & 0.9$\times$0.1@11 
%                  &  -          & -            & -   \\
    NGC2782-core & 1.4$\pm$0.1  & 4.6$\pm$0.6  & 0.2$\times$0.1@20
                  & 0.7$\pm$0.2  & 2.5$\pm$0.7  & 0.1$\times$0.06@153 \\
    NGC2782-jet  & 0.3$\pm$0.1  & 1.8$\pm$1.0  & 0.3$\times$0.2@74
                  & 0.6$\pm$0.2  & 1.4$\pm$0.5  & 0.08$\times$0.06@145  \\
    NGC3147-core & 7.4$\pm$0.2  & 8.6$\pm$0.3  & 0.06$\times$0.04@150
                  & 10.2$\pm$0.1 & 10.0$\pm$0.1 & 0.03$\times$0.03\\
    NGC3718-core & 1.2$\pm$0.1  & 4.0$\pm$0.6  & 0.4$\times$0.1@150 
                  & 5.3$\pm$0.1  & 6.1$\pm$0.3  & 0.05$\times$0.03@139 \\
    NGC3718-jet  & 4.7$\pm$0.2  & 4.5$\pm$0.3  & 0.05$\times$0.03@70 
                  & - & - & -\\
    NGC4579-core & 5.0$^a$ & - & - 
                  & 15.2$\pm$0.2 & 17.2$\pm$0.3 & 0.09$\times$0.03@146 \\
    NGC5953-core & 1.6$\pm$0.2  & 1.5$\pm$0.3  & 0.03$\times$0.03 
                  & 0.5$\pm$0.1  & 0.4$\pm$0.2  & 0.05$\times$0.03@67 \\
    NGC5953-jet  & 1.2$\pm$0.2 & 1.2$\pm$0.3 & 0.3$\times$0.2@10
                  & - & - & - \\
    NGC7217-core & 1.5$\pm$0.1 & 2.6$\pm$0.3 & 0.2$\times$0.1@157 
                  & 5.0$\pm$0.5 & 5.0$\pm$0.5 & 0.04$\times$0.03@13 \\
    \hline
             &\multicolumn {6}{c}{\hrulefill\hskip 3mm
    \raisebox{-1.0mm}{EVN/VLBI}\hskip 3mm \hrulefill} \\                
    & \multicolumn{3}{c}{18cm}
    & \multicolumn{3}{c}{6cm} \\
    Name          & Peak & Integrated & Deconvolved 
                  & Peak & Integrated & Deconvolved  \\
                  & flux & flux density & Size   
                  & flux & flux density & Size \\
                  & [mJy/beam] & [mJy] & $''\times''@\degree$  
                  & [mJy/beam] & [mJy] & $''\times''@\degree$  \\
    \hline
    NGC1961-core & 0.50$\pm$0.07 & 1.2$\pm$0.2  & 0.05$\times$0.01@21  
                  & 0.73$\pm$0.05 & 0.9$\pm$0.1  & 0.004$\times$0.003@55 \\
    NGC2782-core & 0.40$\pm$0.06 & 1.2$\pm$0.2  & 0.05$\times$0.01@170 
                  & $\leq$0.4     &  -           & - \\
    NGC3147-core & 6.1$\pm$0.1   & 6.0$\pm$0.2  & 0.004$\times$0.004
                  & 10.1$\pm$0.3  & 9.3$\pm$0.5  & 0.006$\times$0.006\\
    NGC3718-core & 4.8$\pm$0.1   & 4.7$\pm$0.3  & 0.008$\times$0.007@173
                  & 5.6$\pm$0.1   & 7.1$\pm$0.2  & 0.006$\times$0.005@115 \\
    NGC4579-core & 17.8$\pm$0.6  & 20.7$\pm$1.0 & 0.02$\times$0.01@4 
                  & - & - & - \\     
    NGC4579 (Ulv01)$^b$     & - & 18.3  & - 
                  & - & 22.8 & - \\ 
    NGC5953-core & $\leq$0.3     & - & - 
                  & $\leq$0.4 & - & - \\
    NGC7217-core & 0.5$\pm$0.1   & 0.5$\pm$0.1  & 0.03$\times$0.02@71
                  & 0.41$\pm$0.06 & 1.2$\pm$0.2  & 0.005$\times$0.003@106 \\ 
    \hline
             &\multicolumn {6}{c}{\hrulefill\hskip 3mm
    \raisebox{-1.0mm}{PdBI}\hskip 3mm \hrulefill} \\                
    & \multicolumn{3}{c}{3mm}
    & \multicolumn{3}{c}{1mm} \\
    Name          & Peak & Integrated & Size 
                  & Peak & Integrated & Size  \\
                  & flux & flux density &    
                  & flux & flux density &  \\
                  & [mJy/beam] & [mJy] & $''\times''@\degree$  
                  & [mJy/beam] & [mJy] & $''\times''@\degree$  \\
    \hline
    NGC1961 & 2.6$\pm$0.4 & 3$\pm$0.3 & 3.5$\times$1.8@57 
             & $\leq$1.5 (3$\sigma$) & - & - \\
    NGC2782 & $\leq$1.0 (3$\sigma$) & - & - 
             & $\leq$3.0 (3$\sigma$) & - & - \\
    NGC3147 & 5.3$\pm$0.3  & 5.0$\pm$0.3 & 1$\times$1
             & 2.8$\pm$0.5  & 4.0$\pm$0.4 & 1$\times$1  \\
    NGC3718 & 10$\pm$2 & 11$\pm$2 & 2$\times$2 
             & 9.5$\pm$0.7 & 14$\pm$0.8 &  1$\times$1    \\
    NGC4579 & 11.3$\pm$0.5 & 11$\pm$0.4 & 0.5$\times$0.5 
             & 12$\pm$2 & 11$\pm$2 & 0.5$\times$0.5     \\     
    NGC5953 & $\leq$2 (3$\sigma$) & - & -
             & $\leq$4 (3$\sigma$) & - & - \\
    NGC7217 & $\leq$1.5 (3$\sigma$) & - & -
             & $\leq$4 (3$\sigma$) & - & - \\ 
    \hline
\end{tabular}
\caption{Results of the radio and mm continuum observations with
  MERLIN, EVN and PdBI. $^a$ from the MERLIN archive (PI:
  N.Nagar). $^b$ taken from VLBA observation carried out by Ulvestad
  \& Ho (2001; $\equiv$Ulv01). Please note that the flux errors do not
  include the calibration uncertainties of 5-10\% and represent only
  the statistical uncertainties of the fit. The given angles are
  derived from North to East. }
\label{radio-fluxes}
\end{table*}

\subsubsection{NGC~1961}
NGC~1961 contains a LINER type 2 nucleus in an SAB(rs)c host galaxy at
a distance of 52~Mpc.  This is the most distant source in our survey.
Assuming $\sigma_{\rm s}$=255\kms (McElroy 1995), $M_{\rm bh}$ is
inferred to be 3.0$\times10^8$\msun\, (Table~\ref{radio-lum}). Taking
$L_{\rm x}\simeq4\times10^{40}$erg~s$^{-1}$ (Roberts \& Warwick 2000),
the nuclear bolometric luminosity of NGC~1961 can be estimated as
$\sim$10$^{41}$erg\,s$^{-1}$ implying a sub-Eddington\footnote{$L_{\rm
ed}$/\lsun$=3.22\times10^4$M/\msun} system with $L_{\rm bol}/L_{\rm
ed}\simeq10^{-6}-10^{-5}$ (Table~\ref{radio-lum}).

\begin{figure*}[!]
\centering
\resizebox{\hsize}{!}{\rotatebox{-90}{\includegraphics{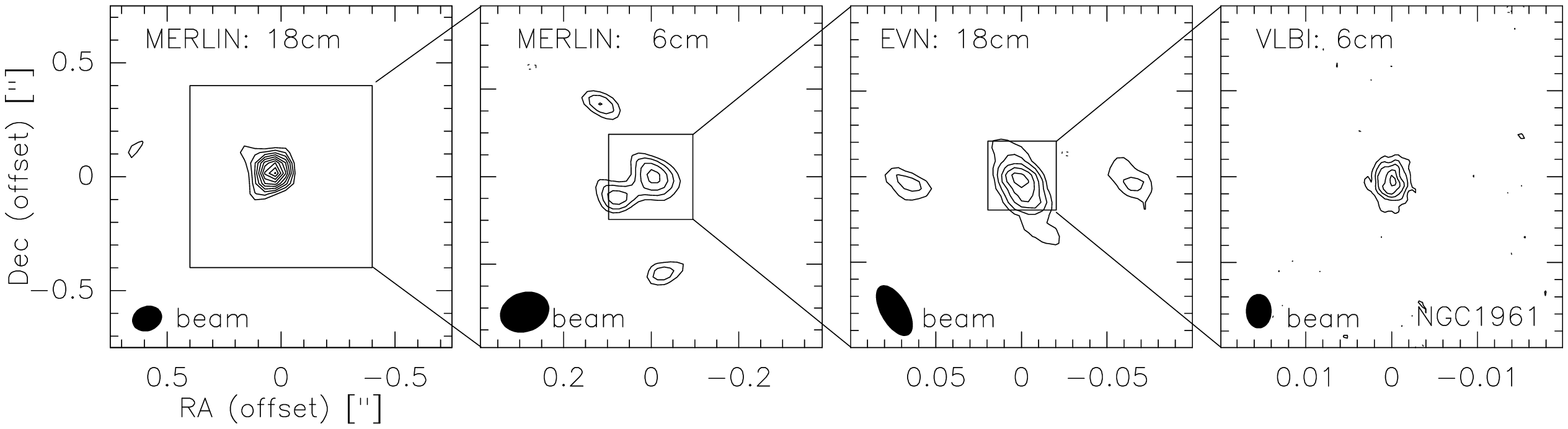}}}
\vskip 0.4cm
\resizebox{\hsize}{!}{\rotatebox{-90}{\includegraphics{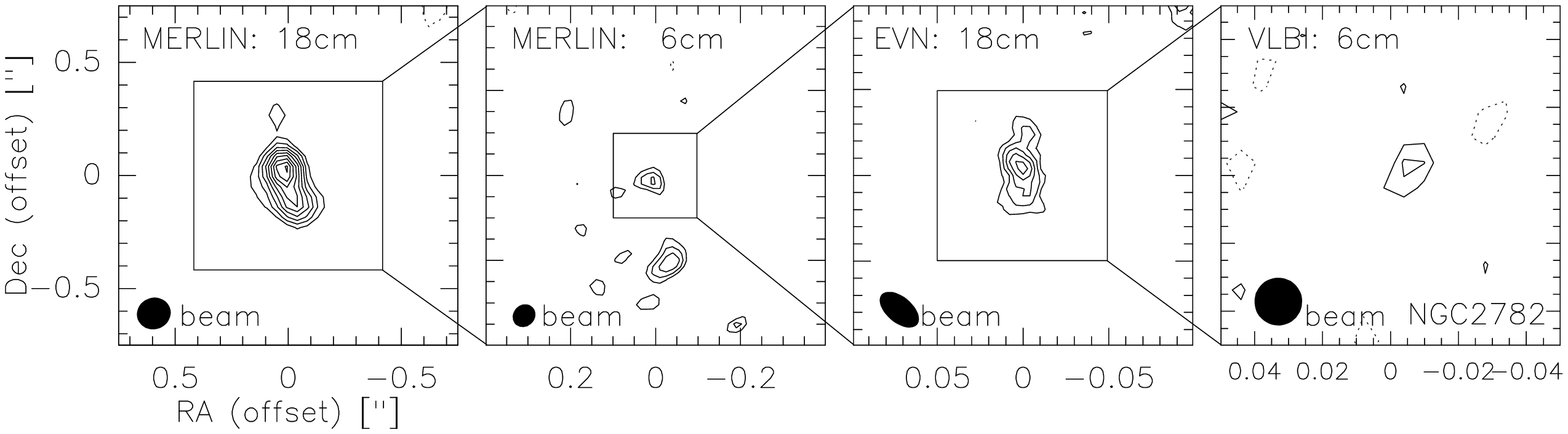}}}
\vskip 0.4cm
\resizebox{\hsize}{!}{\rotatebox{-90}{\includegraphics{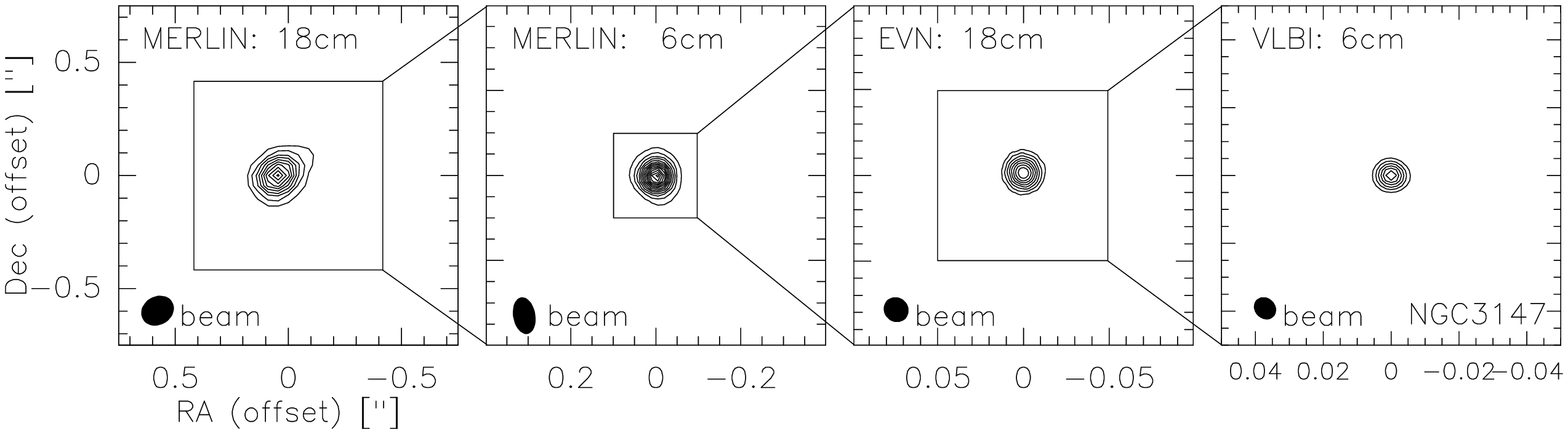}}}
\caption{MERLIN (2 {\it left} columns) and EVN/VLBI (two {\it right}
  columns) of the 18cm and 6cm continuum emission in NGC1961 ({\it
  upper} panels), NGC2782 ({\it middle} panels) and NGC3147 ({\it
  lower} panels).  The respective synthesized beams are shown in the
  lower left part of each map. Contour levels for MERLIN-18cm
  (NGC~1961; {\it upper left}): (3$\sigma$=)0.57 to 2.28\,mJy/beam in
  steps of 1$\sigma$; for MERLIN-6cm ({\it upper left middle}):
  (3$\sigma$=)0.39 to 0.78\,mJy/beam in steps of 1$\sigma$; EVN-18cm
  ({\it upper right middle}): (3$\sigma$=)0.21 to 0.49\,mJy/beam in
  steps of 1$\sigma$; EVN-6cm ({\it upper right}): (3$\sigma$=)0.14 to
  0.61\,mJy/beam in steps of 1$\sigma$; MERLIN-18cm (NGC2782; {\it
  middle left}): (3$\sigma$=)0.45 to 1.7\,mJy/beam in steps of
  1$\sigma$, for MERLIN-6cm ({\it middle left middle}):
  (3$\sigma$=)0.39 to 0.78\,mJy/beam in steps of 1$\sigma$, for
  EVN-18cm ({\it middle right middle}): (3$\sigma$=)0.18 to
  0.36\,mJy/beam in steps of 1$\sigma$; for EVN-6cm ({\it middle
  left}): (2$\sigma$=)0.2 to 0.3\,mJy/beam in steps of
  1$\sigma$. MERLIN-18cm (NGC3147; {\it lower left}): (5$\sigma$=)0.75
  to 7.5\,mJy/beam in steps of 5$\sigma$, for MERLIN-6cm ({\it lower
  left middle}): (3$\sigma$=)0.72 to 10.1\,mJy/beam in steps of
  5$\sigma$, for EVN-18cm ({\it lower right middle}): (3$\sigma$=)0.65
  to 5.2\,mJy/beam in steps of 5$\sigma$; for EVN-6cm ({\it lower
  left}): (5$\sigma$=)1.4 to 8.3\,mJy/beam in steps of
  5$\sigma$. Negative contours, if visible, correspond to 3$\sigma$.}
\label{1st-3-radio}
\end{figure*}

The MERLIN 18cm map clearly shows extended emission, consistent with
the difference between peak and integrated flux densities (see
Fig.~\ref{1st-3-radio} and Table~\ref{radio-fluxes}) and previous VLA
maps which also indicate extended emission (Condon, Anderson \&
Broderick 1995). The deconvolved size is slightly bigger than the beam
size, giving further evidence for extended emission. A two-component
Gaussian fit assuming a core and a ``jet-like'' feature results in a
10$\sigma$ determination of the core flux but only in a
$\sim$2$\sigma$ estimate for the flux of the potential jet. The MERLIN
6cm map independently indicates extended emission and, besides a
compact core, three additional components. The northeastern and the
southeastern features lie in the direction of the extensions seen in
the MERLIN 18cm map. However, given that the southwestern feature
appears to be symmetrically positioned on the other side of the core
with respect to the northeastern feature, sidelobe effects cannot be
totally excluded for these two components. Also, since the
signal-to-noise ratios of all three ``off-core'' components are still
marginal, they remain questionable. We find indications of extended
emission in the EVN 18cm data as well by comparing the peak with the
integrated flux density. The distribution appears to be slightly
resolved and shows an elongation to the northeast but with a different
PA than found in the MERLIN maps. The extensions are in the direction
of the beam, so we must again consider sidelobe effects.  The two
artificial components to the west and to the east are found to be at
the positions of the sidelobes. The 6cm EVN component is only
marginally resolved and the deconvolved size is smaller than the beam
size (by a factor of $\sim$2). The PdBI 3mm map also reveals a
jet-like component of $\sim$1mJy which is $\sim$4$''$ northeast of the
nucleus (Baker et al.\, in prep.). Further radio observations are
needed to clarify the different orientations of the jet-like
components found at different wavelengths and angular resolution.

Both EVN and MERLIN data imply a spectral index of $\alpha = -0.3$
(for $f_\nu \propto \nu^\alpha$), indicating a decrease of flux with
decreasing wavelength. The higher resolutions of the 6\,cm maps mean
that the spectral index might be higher and so the spectrum even
flatter.  The upper limit on the spectral index derived from the PdBI
fluxes is similar to the cm-derived values and also suggests a flat
spectrum.

\begin{figure*}[!]
\centering
\resizebox{\hsize}{!}{\rotatebox{-90}{\includegraphics{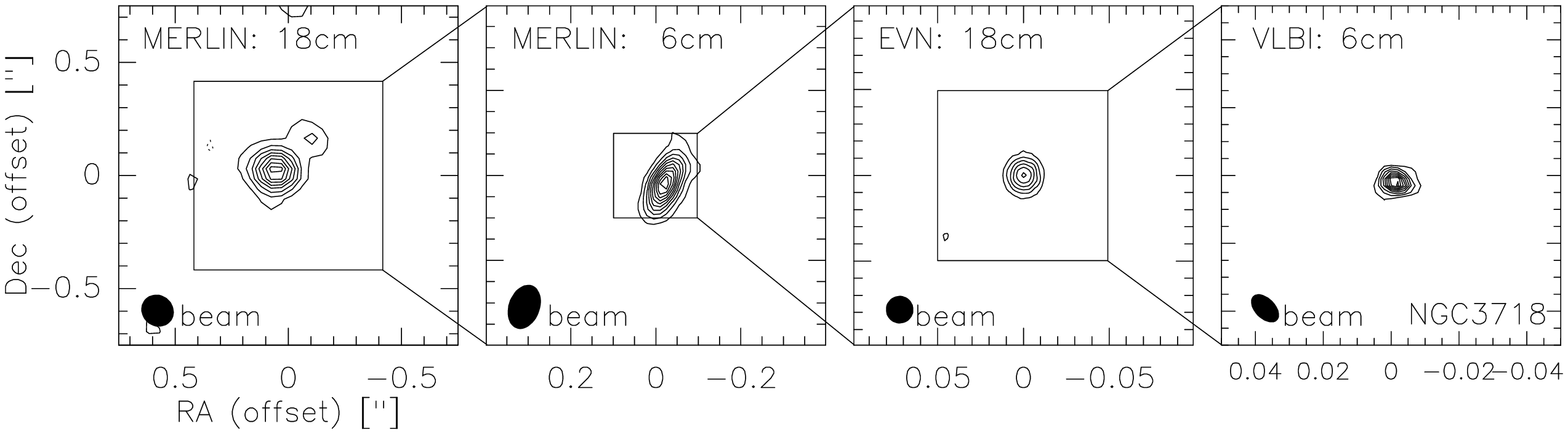}}}
\vskip 0.4cm
\resizebox{\hsize}{!}{\rotatebox{-90}{\includegraphics{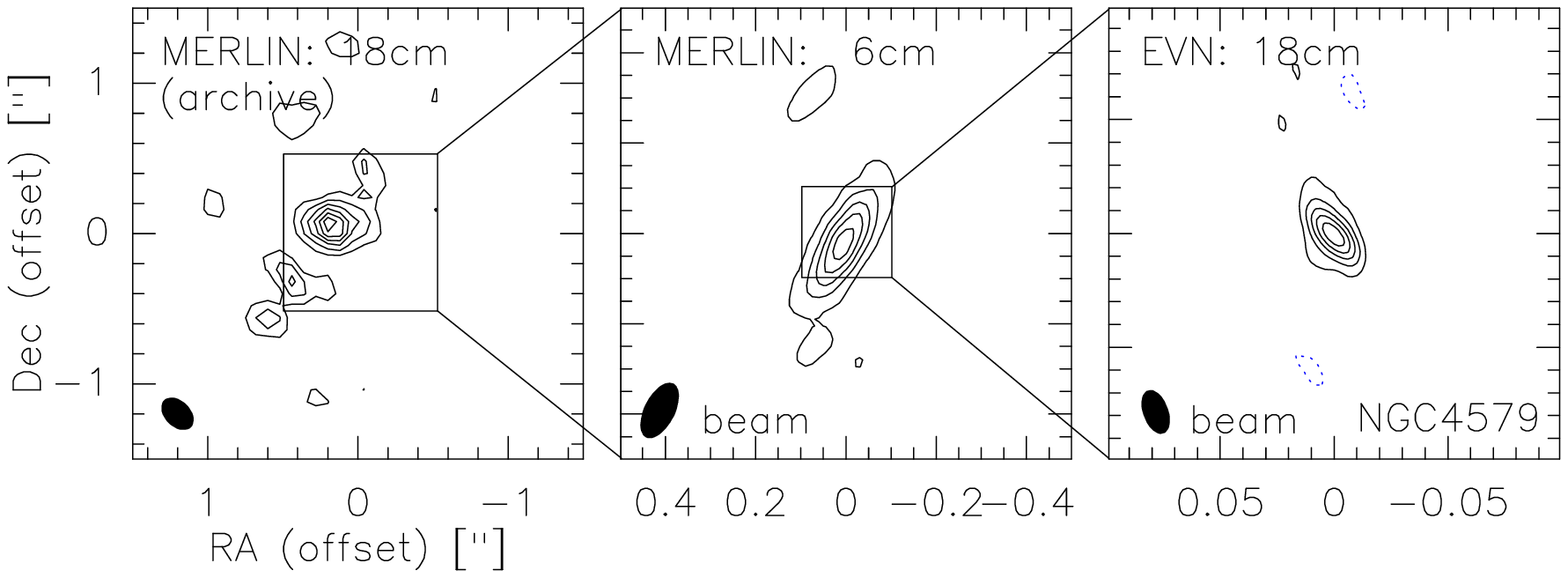}}}
\vskip 0.4cm
\resizebox{\hsize}{!}{\rotatebox{-90}{\includegraphics{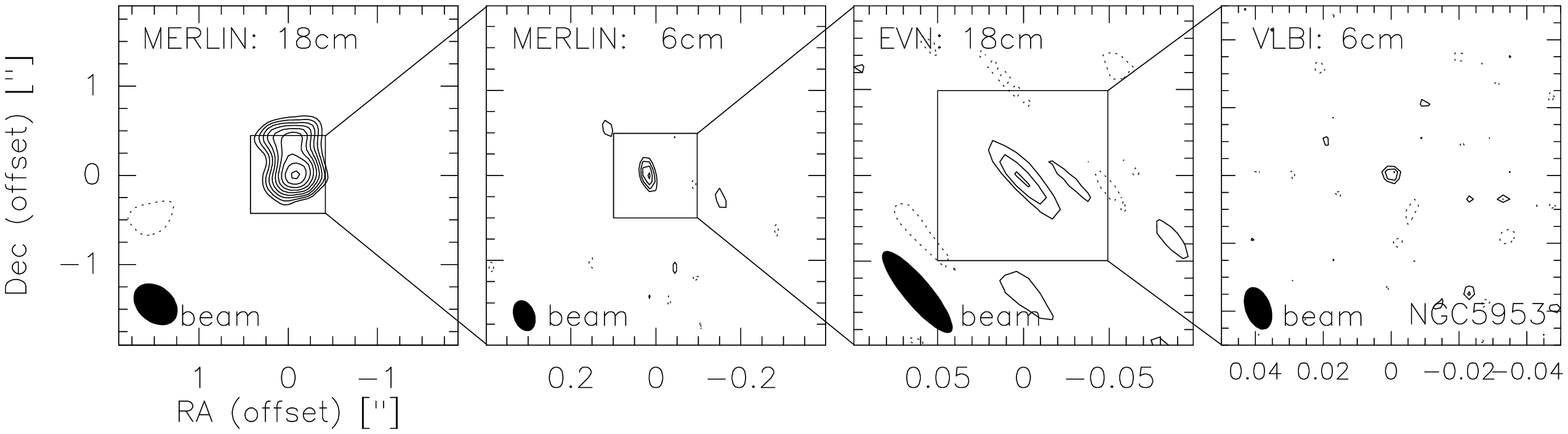}}}
\caption{MERLIN (2 images on the {\it left} respectively) and EVN (2
  images on the {\it right} respectively) of the 18cm and 6cm
  continuum emission in NGC3718 ({\it upper} panel), NGC4579 ({\it
  middle} panel) and NGC5953 ({\it lower} panel) . The respective
  synthesized beams are shown in the lower left corner of each map.
  MERLIN-18cm (NGC~3718; {\it upper left}): (5$\sigma$=)0.64 to
  5.1\,mJy/beam in steps of 5$\sigma$, for MERLIN-6cm ({\it upper left
  middle}): (5$\sigma$=)0.5 to 5.0\,mJy/beam in steps of 5$\sigma$,
  for EVN-18cm ({\it upper right middle}): (5$\sigma$=)0.75 to
  4.5\,mJy/beam in steps of 5$\sigma$; for EVN-6cm ({\it upper
  right}): (5$\sigma$=)0.55 to 5.0\,mJy/beam in steps of 5$\sigma$;
  Contour levels for MERLIN-6cm (NGC4579; {\it middle left middle}):
  (5$\sigma$=)1.0 to 15.0\,mJy/beam in steps of 10$\sigma$, for
  EVN-18cm ({\it middle right middle}): (3$\sigma$=)3.3 to
  16.5\,mJy/beam in steps of 3$\sigma$; MERLIN-18cm (NGC5953; {\it
  lower left}): (3$\sigma$=)0.48 to 1.5\,mJy/beam in steps of
  1$\sigma$, for MERLIN-6cm ({\it lower left middle}):
  (3$\sigma$=)0.74 to 1.7\,mJy/beam in steps of 1$\sigma$, for
  EVN-18cm ({\it lower right middle}): (2$\sigma$=)0.12 to
  0.24\,mJy/beam in steps of 1$\sigma$; for EVN-6cm ({\it lower
  right}): (2$\sigma$=)0.2 to 0.4\,mJy/beam in steps of 1$\sigma$.
  Negative contours, if visible, correspond to 3$\sigma$.}
\label{2nd-3-radio}
\end{figure*}

\begin{figure*}[!]
\centering
\resizebox{\hsize}{!}{\rotatebox{-90}{\includegraphics{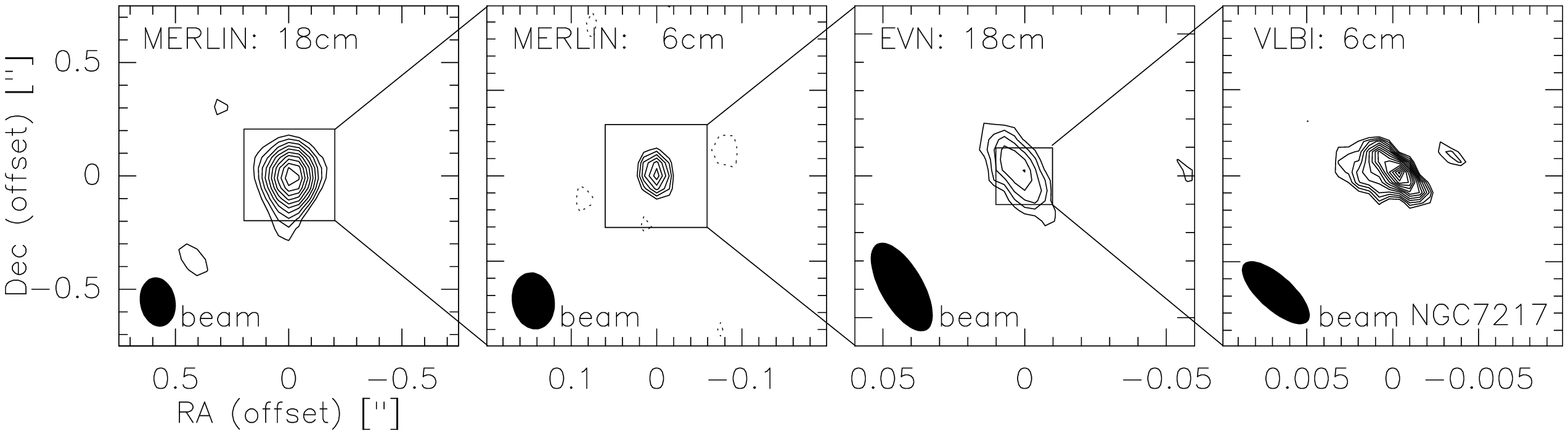}}}
\caption{MERLIN (2 images on the {\it left}) and EVN (2 images on the
  {\it right}) of the 18cm and 6cm continuum emission in NGC7217. The
  respective synthesized beams are shown in the lower middle part of
  each map. Contour levels for MERLIN-18cm ({\it left}):
  (3$\sigma$=)0.36 to 1.4\,mJy/beam in steps of 1$\sigma$, for
  MERLIN-6cm ({\it left middle}): (3$\sigma$=)1.5 to 4.0\,mJy/beam in
  steps of 1$\sigma$, for EVN-18cm ({\it right middle}):
  (3$\sigma$=)0.21 to 0.5\,mJy/beam in steps of 1$\sigma$; for EVN-6cm
  ({\it right}): (3$\sigma$=)0.2 to 0.6\,mJy/beam in steps of
  1$\sigma$. Negative contours, if visible, correspond to 3$\sigma$.}
\label{NGC7217-radio}
\end{figure*}

\subsubsection{NGC~2782}
The SAB(rs)pec host galaxy of NGC~2782 (D=35~Mpc) harbours a dominant
nuclear HII region; historically, its nuclear activity has been
supposed to be mainly dominated by a powerful nuclear starburst rather
than by an AGN.  However, recent observations of water maser emission
in this object suggest that a hidden AGN might be present as well
(Braatz et al.\ 2004). Its central black hole mass can be estimated as
4$\times10^7$\msun\, assuming $\sigma_{\rm s}$=146\kms (McElroy
1995). The nuclear bolometric luminosity amounts to
$\sim$10$^{41}$erg~s$^{-1}$ assuming $L_X$ from
Table~\ref{radio-lum}. A comparison with the Eddington luminosity
indicates that NGC~2782 might also radiate at a sub-Eddington rate
with $L_{\rm bol}/L_{\rm ed}\simeq10^{-5}-10^{-4}$.

A strong core and a strong extended jet are visible in both the MERLIN
18cm and 6cm maps (see Fig.~\ref{1st-3-radio} and
Table~\ref{radio-fluxes}), supporting the presence of an optically
hidden AGN; VLA maps also show extended and diffuse emission on
arcsecond scales (Saikia et al.\ 1994). The jet is elongated towards
the south/southeast and has a length of $\sim$0.3$''$ (= 53~pc). The
core seems to be slightly extended, too. Going to 6cm and thus to
higher angular resolution, the core and the jet become more resolved,
as indicated also by their lower fluxes.  The positions of both
components agree at the two wavelengths. The EVN 18cm maps also show
an extended component, again consistent with the different peak and
integrated flux densities. The extension is not in the direction of
the beam. However, the PA of the extended emission is slightly
different from that in the MERLIN maps, raising doubts about its
reliability.  In the EVN 6cm map, we find a $\sim$3$\sigma$ peak at
the position of the nucleus. Resolution effects may play a role for
this tentative detection. The spectral index derived from the MERLIN
data agrees with the upper limit obtained from the EVN data. Parts of
the jets might affect the core emission in the MERLIN maps since core
and jet emission cannot be easily separated from each other in the
18cm MERLIN map, complicating the estimate of the spectral index for
the core and the jet.

No continuum emission was detected at 3~mm or 1~mm with the IRAM PdBI.
The derived upper limits are given in Table~\ref{radio-fluxes}.

\subsubsection{NGC~3147}
NGC~3147 is classified as SA(rs)bc galaxy at a distance of 38~Mpc with
a Seyfert type 2 nucleus. The central black hole mass can be derived
as $\sim$4$\times$10$^8$\msun\, with $\sigma_{\rm s}=268$\kms (McElroy
et al.\ 1995) and the nuclear bolometric luminosity is
$\sim$2$\times$10$^{42}$erg~s$^{-1}$ (Table~\ref{radio-lum}), implying
$L_{\rm bol}/L_{\rm ed}\simeq5\times10^{-5}$. Therefore, NGC~3147
radiates at a sub-Eddington rate.

\begin{figure}[!]
\centering
\resizebox{\hsize}{!}{\rotatebox{-90}{\includegraphics{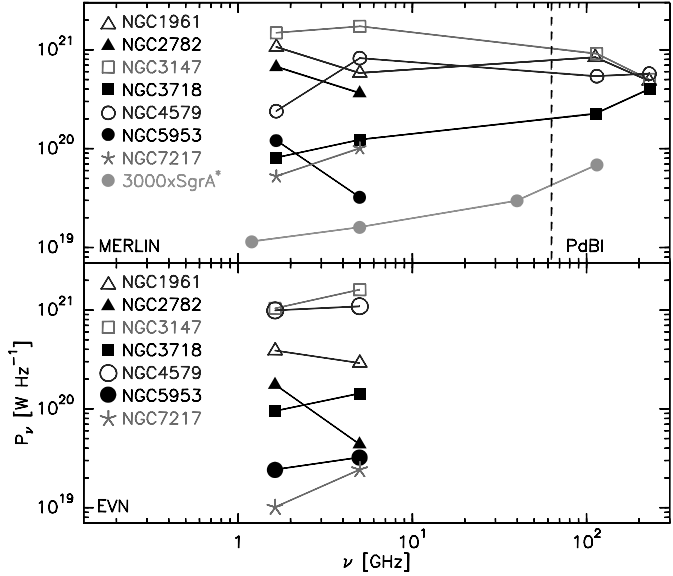}}}
\caption{Spectral energy distribution from the MERLIN/PdBI ({\it upper
  panel}) and EVN data ({\it lower panel}). The dashed line in the
  upper panel separates the MERLIN from the PdBI data. Values for
  SgrA$^\star$ were taken from Falcke et al.\ 1998 and Zhao et al.\
  2001.}
\label{specind}
\end{figure}

NGC~3147 is among the strongest radio sources in our survey (see
Fig.~\ref{1st-3-radio} and Table~\ref{radio-fluxes}).  In all maps
(including the mm-data and VLA data (Ho \& Ulvestad 2001), it appears
to be point-like. The almost identical fluxes determined at both
wavelengths and instruments are a further sign of very compact
emission in this object. The flat to inverted spectral index derived
from the EVN/MERLIN flux densities indicates that no extended jet is
present. The point-like structure and the spectral indices (at
cm-wavelengths) are consistent with what Ulvestad \& Ho (2002) found
in VLBA data (see also Anderson, Ulvestad \& Ho 2004; Anderson \&
Ulvestad 2005). Our two observing epochs for the EVN 18cm data show no
evidence for any variability of this source.  We find a turnover
between the cm- and the mm-fluxes (Fig.~\ref{specind}), as can also be
seen in the different spectral indices at mm- and cm-wavelengths
(Table~\ref{radio-specind}). The turnover is most likely caused by
synchrotron self absorption, as is discussed in further detail in a
separate paper (Krips et al. 2006).

\subsubsection{NGC~3718}
The galaxy NGC~3718 is located at a distance of 13~Mpc, hosts a LINER
type 1.9 nucleus, and has a warped gas and dust disk which might be
caused by an interaction with a close companion galaxy (e.g., Krips et
al.\ 2005). A black hole mass of $\sim$5$\times$10$^7$\msun\, is
implied by $\sigma_{\rm s}$ = 157\kms (Ho, Filippenko, \& Sargent
1997), and a nuclear bolometric luminosity of
$\sim$10$^{41}$erg~s$^{-1}$ (Table~\ref{radio-lum}) implies a
sub-Eddington system with $L_{\rm bol}/L_{\rm
ed}\simeq10^{-5}-10^{-4}$.

Similar to NGC~3147, NGC~3718 is among the strongest radio continuum
sources in our survey (Table~\ref{radio-fluxes}). It shows extended
emission on MERLIN 18cm scales and signs of a quite compact
($\sim$0.5$'' = $34~pc) $4\sigma$ jet in a northwest direction
(Fig.~\ref{2nd-3-radio}). This is supported by VLA maps that also
reveal extended emission on large scales (Condon 1987). There are some
weak indications for extended emission from inspection of the 6cm maps
and comparison of the peak and integrated flux densities. Moreover,
the EVN 6cm map reveals a slight extension but, in contrast to the
MERLIN 18cm map, in a westward direction, which we therefore do not
identify as the same jet. The flux densities from MERLIN and EVN are
identical, indicating that there is no emission extended on
intermediate scales in NGC\,3718 other than the small-scale jet. Our
fluxes are in good agreement with those obtained in VLBA observations
by Nagar et al.\ (2002). The spectral indices are very consistent for
the MERLIN and EVN data and are derived to be 0.4, describing an
inverted spectrum.

\subsubsection{NGC~4579}
NGC~4579 is a SAB(rs)b galaxy at a distance of 20~Mpc with a Seyfert
type 1.9 and/or LINER type 1.9 nucleus. The black hole mass is
$\sim$10$^8$\msun\, assuming $\sigma_{\rm s}$=185\kms (McElroy
1995). Ho et al.\ (1999) determine the nuclear bolometric luminosity
to be $\sim$10$^{42}$erg~s$^{-1}$. This gives a value of
$\sim$10$^{-4}$ for $L_{\rm bol}/L_{\rm ed}$ indicating a
sub-Eddington system.

NGC~4579 was already observed with MERLIN at 18cm (PI: N.Nagar, from
the MERLIN archive) and with the VLBA at 18cm, 13cm, 6cm and 4cm
(Ulvestad \& Ho 2001). We thus conducted only complementary
observations with MERLIN at 6cm and EVN at 18cm (Fig.\ref{2nd-3-radio}
and Table~\ref{radio-fluxes}).  In the VLBA maps from Ulvestad \& Ho
(2001), no extended emission can be seen, while VLA maps from Ho \&
Ulvestad (2001) at 20cm and 6cm show a jet-like extension of
~$\sim$3$''$ to the northwest. Our 18~cm and 6~cm maps do not show any
jet-like feature on sub-arcsecond scales, but the difference between
peak flux and integrated flux still indicates extended
emission. Furthermore, the deconvolved size at 18cm is slightly larger
than the beam, supporting the presence of extended emission. Falcke et
al.\ (2001) report variability of the radio flux at 15~GHz on
timescales of 1-3\,years. We too find a difference of $\sim 10\%$
between the EVN and VLBA 18\,cm flux densities.  Thus, the
determination of the spectral index must be made and interpreted with
caution since not all fluxes were measured simultaneously. This might
explain the large discrepancy between the MERLIN 18cm flux and that
from the EVN 18cm data, which is {\it higher} by a factor of 2-3.

\subsubsection{NGC~5953}
NGC~5953 is characterised as a SAa pec galaxy at a distance of 26~Mpc
hosting a Seyfert type 2 nucleus. It is also known to interact with a
close companion NGC~5954 (e.g., Iono, Yu \& Ho 2005). The black hole
mass of $\sim$7$\times$10$^6$\msun\, ($\sigma_{\rm s}$=94\kms; McElroy
1995) is the smallest one in our sample.  Woo \& Urry (2002) find a
surprisingly high nuclear bolometric luminosity of
$\sim$10$^{44}$erg~s$^{-1}$ but their estimate is based on fluxes
taken over very large apertures. No X-ray luminosity has been
published yet. Thus, any conclusion about its accretion efficiency
remains speculative.

NGC~5953 turns out to be the weakest radio source in our sample
(Table~\ref{radio-fluxes}). It is clearly detected in the MERLIN 18cm
and 6\,cm maps, while it remains undetected in the EVN maps
(Fig.~\ref{2nd-3-radio}). We find 4$\sigma$ peaks in the latter at the
position of the nucleus. The MERLIN 18cm data show a strong jet to the
north, which is slightly resolved with the MERLIN beam
($\sim$0.3$''\equiv$41pc). A steep spectral index for the core is
derived from the MERLIN data, which might be biased by resolution
effects in the MERLIN 6cm map. VLA maps (e.g., Condon et al.\ 1990;
Iono, Yu \& Ho 2005) also indicate extended emission on large angular
scales, but the emission is somewhat smeared between the two
interacting systems in the VLA beam.

\begin{table}
\centering
\begin{tabular}[c]{ccccc}
    \hline
    \hline
    Name     & $\alpha^{230}_{113}$ & $\alpha^{5}_{1.7}$
    & $\alpha^{5}_{1.6}$ & $\alpha^{5}_{1.7}$ \\
             & (PdBI) &  (MERLIN) & (EVN) & (VLBA)$^a$ \\
    \hline
    NGC1961 & $-$0.4  & $-$0.3$\pm$0.3   & $-$0.3$\pm$0.2   & - \\
    NGC2782 & -       & $-$0.6$\pm$0.3   & $\leq-$1.0     & - \\
        jet  &  -      & $-$0.2$\pm$0.6   & -              & - \\
    NGC3147 & $-$0.3  &  0.14$\pm$0.03 &  0.39$\pm$0.06 & 0.2$\pm$0.1 \\
    NGC3718 & 0.0$\pm0.2$  &  0.4$\pm$0.1   &  0.37$\pm$0.06 & $-$ \\
    NGC4579 & 0.1  &  1.12$\pm$0.02 &  0.09$\pm$0.04 & 0.2$\pm$0.1 \\
    NGC5953 & -     & $-$1.2$\pm$0.5   &   -            & - \\
    NGC7217 & -     &  0.6$\pm$0.1   & 0.8$\pm$0.2    &  -  \\
    \hline
\end{tabular}
\caption{Spectral indices of the respective sources.
 $S_\nu\propto\nu^\alpha$. $^a$ taken from Ulvestad \& Ho 2001.}
\label{radio-specind}
\end{table}

\begin{table}
\centering
\begin{tabular}[c]{ccccc}
    \hline
    \hline
    Name & $\nu$ & S$_\nu$ & $\theta$ & T$_b$ \\
         & [GHz] & [mJy]   & [mas]    & [K]   \\
    \hline
    NGC1961     & 5.0 &  0.9 & $\leq$1  & $\geq$4$\times$10$^7$\\
    NGC2782     & 1.6 &  1.2 & $\leq$10 & $\geq$2$\times$10$^6$\\
    NGC3147     & 5.0 & 10.1 & $\leq$4  & $\geq$3$\times$10$^7$\\
    NGC3718     & 5.0 &  7.1 & $\leq$1  & $\geq$3$\times$10$^8$\\
    NGC4579$^a$ & 5.0 & 22.8 & $\leq$4  & $\geq$7$\times$10$^8$\\
    NGC5953     & 5.0 &  0.5 & $\leq$50 & $\geq$1$\times$10$^4$\\
    NGC7217     & 5.0 &  1.2 & $\leq$3  & $\gtrsim$7$\times$10$^6$\\
    \hline
\end{tabular}
\caption{Brightness temperatures estimated with the formula given in
Condon et al.\ (1982). The frequency ($\nu$), the flux density
(S$_\nu$) and the deconvolved size ($\theta$) as an upper limit are
taken from Table~\ref{radio-fluxes}. $^a$ taken from Ulvestad \& Ho
2001.}
\label{temp-b}
\end{table}

\subsubsection{NGC~7217}
NGC~7217 contains a LINER type 2 nucleus in an (R)SAB(rs)a host galaxy
at a distance of 13~Mpc. It has a black hole mass of
$\sim$3$\times$10$^7$\msun\, assuming $\sigma_s=132$\kms (McElroy
1995) and an estimated nuclear bolometric luminosity of
$\sim$10$^{40}$erg~s$^{-1}$ (Table~\ref{radio-lum}). Both values
result in $L_{\rm bol}/L_{\rm ed}\simeq10^{-6}-10^{-5}$. Thus,
NGC~7217 is also a sub-Eddington system.

The MERLIN 18cm map (Fig.~\ref{NGC7217-radio} and
Table~\ref{radio-fluxes}) clearly shows an extended component with an
elongation to the south, although NGC~7217 appears to be unresolved on
VLA scales.  The MERLIN 6cm and EVN 18cm component remains pointlike,
however, while the EVN 6cm map reveals an extended component but in an
eastern direction different from that in the MERLIN 18cm map. This
apparent inconsistency might be a result of the different sensitivity
levels and angular resolutions obtained by the respective
observations. The MERLIN and EVN spectra are the most strongly
inverted in our sample.

\begin{figure}[!]
\centering
\resizebox{8cm}{!}{\rotatebox{-90}{\includegraphics{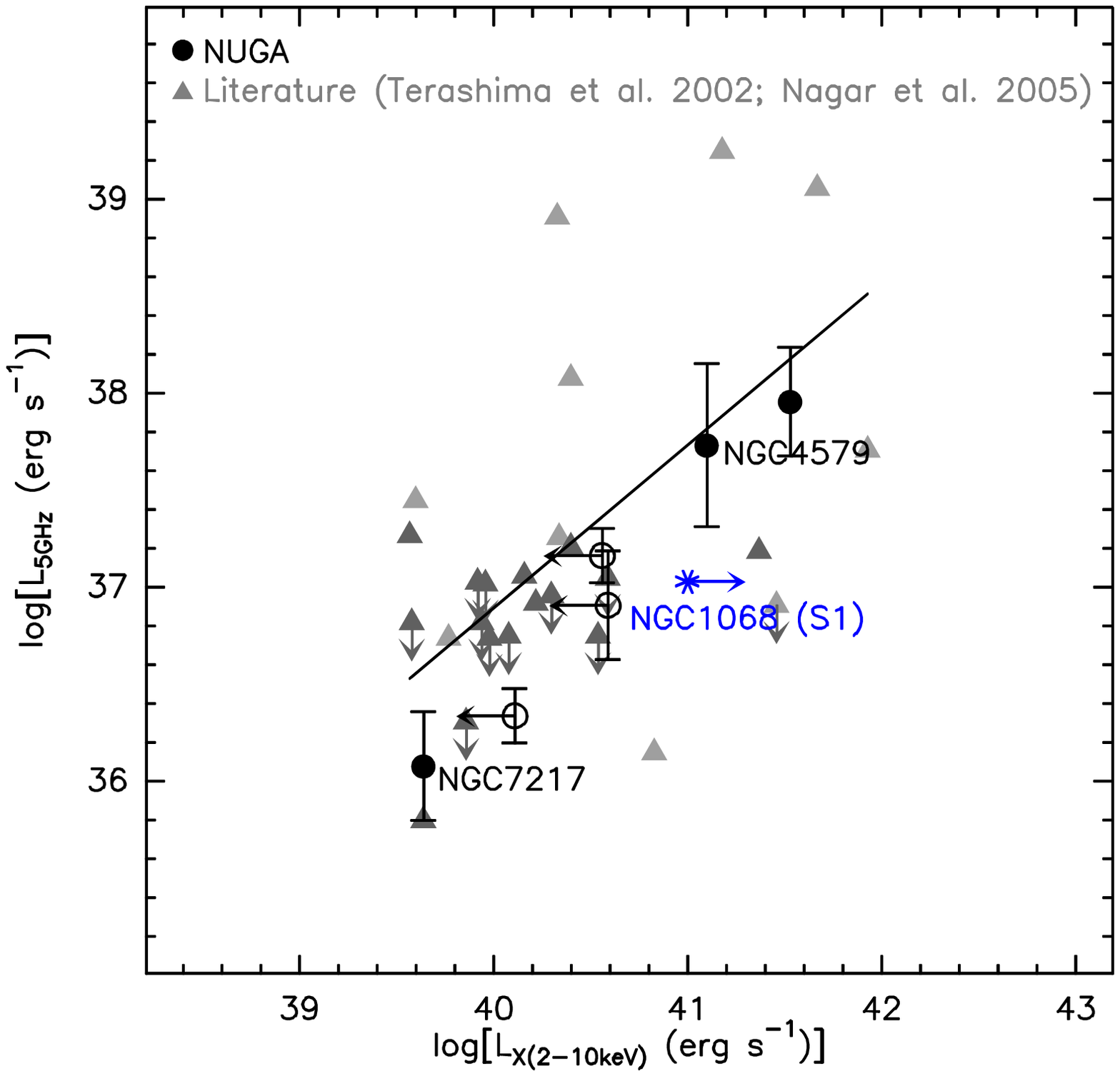}}}
\resizebox{8cm}{!}{\rotatebox{-90}{\includegraphics{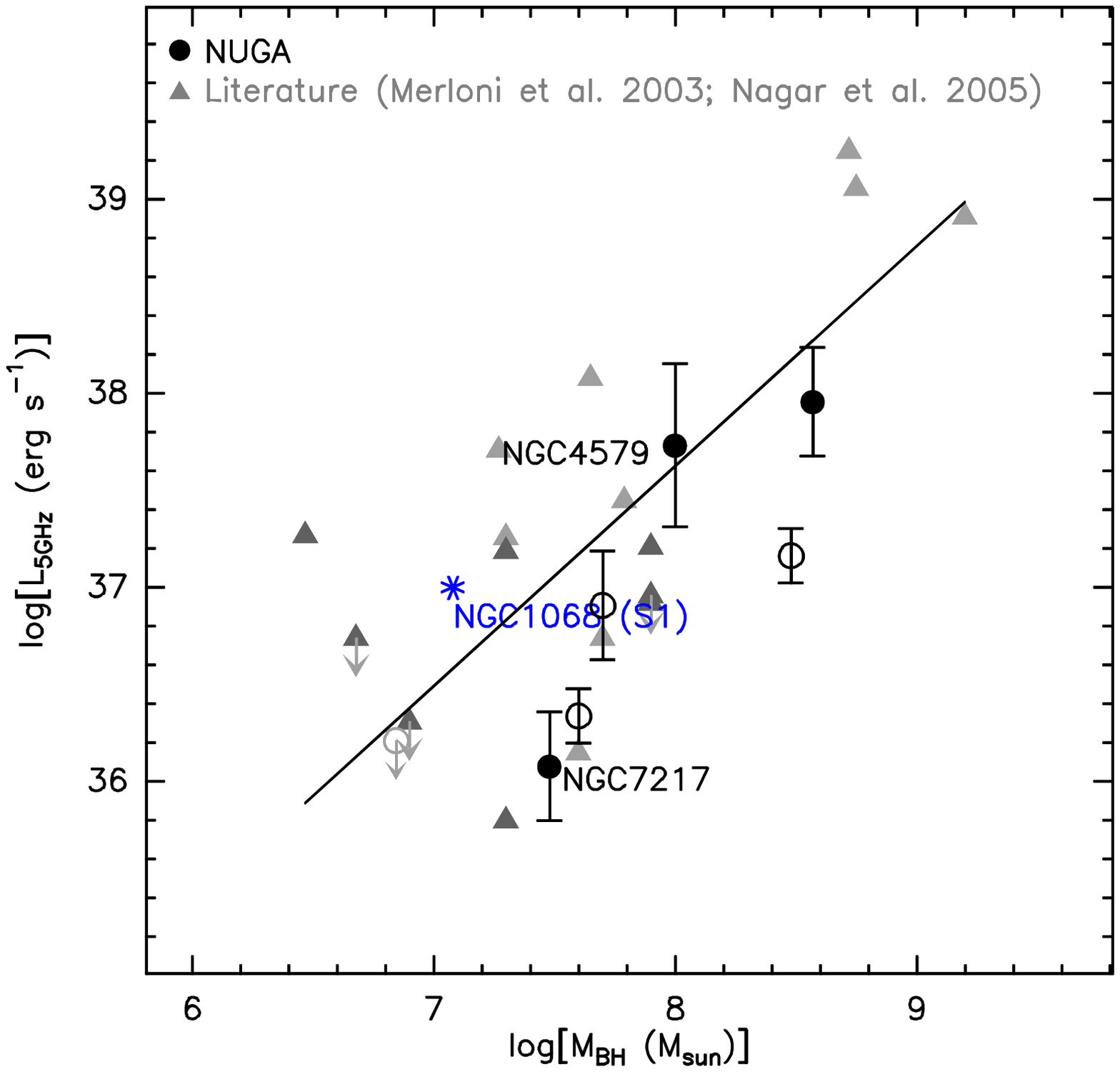}}}
\caption{Comparison of radio (5~GHz) with X-ray luminosities ({\it
  upper panel}) and with black hole mass ({\it lower panel:}). We
  extrapolated the 15~GHz literature data ({\it triangles}) to 5~GHz
  where necessary using the spectral indices indicated in Merloni,
  Heinz \& Di~Matteo (2003); this might cause a larger scatter of the
  data because of their larger uncertainty. The {\it filled light grey
  triangles} represent VLA data while the {\it filled dark grey
  triangles} are taken with the VLBA. For those galaxies from our
  sample, which lack published hard (2-10keV) X-ray luminosities ({\it
  open black circles}), we assumed L$_{\rm
  X}$(0.2-4keV)$\gtrsim$L$_{\rm X}$(2-10keV) as simplification (see
  text for discussion). The {\it open grey circle} represents
  NGC~5953. The literature values are taken from Terashima et al.\
  (2002; X-ray), Nagar, Falcke \& Wilson (2005; VLBA or VLA data at
  15~GHz) and Merloni, Heinz \& Di~Matteo et al.\ (2003; M$_{\rm
  BH}$). Errors of the radio luminosities are from the EVN 5.0~GHz
  data. The radio data ({\it blue star}) from NGC~1068 refer to
  component S1 which is supposed to be the AGN and are taken from
  Gallimore et al.\ (2004), while the black hole mass is from Hur\'e
  (2002) and Lodato \& Bertin (2003). The X-ray luminosity is adopted
  from Merloni, Heinz \& Di~Matteo (2003) but has to be regarded as a
  lower limit due to a possible heavy obscuration of the AGN in
  NGC~1068.}
\label{lum-mbh}
\end{figure}

\begin{figure}[!]
\centering
\resizebox{\hsize}{!}{\rotatebox{-90}{\includegraphics{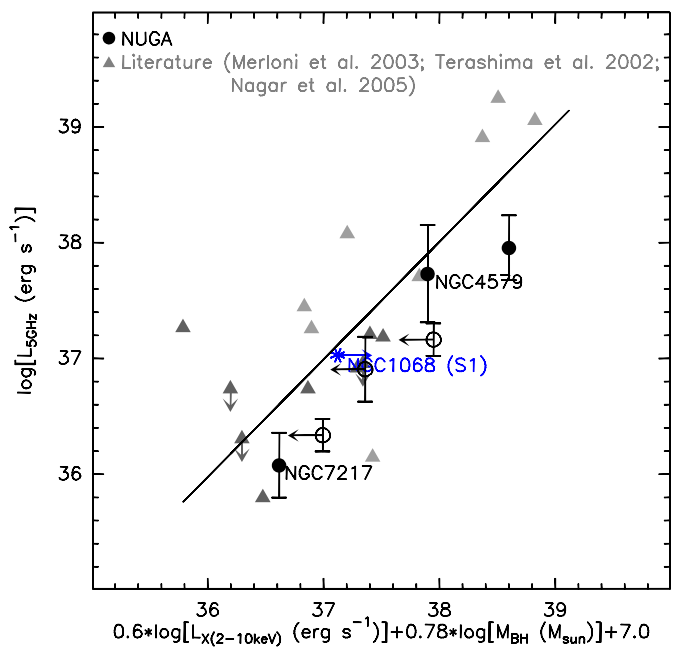}}}
\caption{Comparison of radio fluxes with a projection of the ``fundamental plane''
derived by Merloni, Heinz \& Di~Matteo (2003). See caption of
Fig.~\ref{lum-mbh} for more details on data.}
\label{lum-merlo}
\end{figure}

\begin{table*}
\centering
\begin{tabular}[c]{lcccccccccc}
\hline
\hline
NGC & D$^l$ & $M_{\rm bh}$ & log($L_{\rm 5.0}^{\rm VLBI}$) & 
log($L_{\rm 5.0}^{\rm VLA}$) &
log($L_{\rm x}$)$^n$ & log($L_{\rm bol}$) & log($L_{\rm ed}$)$^i$
& log($L_{\rm bol}$/$L_{\rm ed}$) & log($L_{\rm 5.0}^{\rm obs}$/$L_{\rm x}$) 
\\
& [Mpc] & [10$^8$\,\msun]   & [erg\,s$^{-1}$]$^j$ & [erg\,s$^{-1}$]$^m$ 
& [erg\,s$^{-1}$] & [erg\,s$^{-1}$] & [erg\,s$^{-1}$] & range & \\
\hline
1961 & 52 & 3.4$^a$     & 37.16 & $<$38.02 & 40.56$^c$ & $\sim$41$^g$ & 46.6
& [$-$5,$-$6] & -4.23   \\
2782 & 35 & 0.4$^a$     & 36.34 & 38.02 &  40.11$^c$ & $\sim$41$^g$ & 45.7
& [$-$4,$-$5] & -3.77  \\
3147 & 38 & 4.2$^a$     & 37.91 & 38.23 &  41.51$^d$ & $\sim$42.4$^g$ & 46.7
& [$-$4,$-$5]  & -3.62 \\
3718 & 13 & 0.5$^{a,b}$ & 36.86 & 37.60 &  40.59$^e$ & $\sim$41$^g$ & 45.8
& [$-$4,$-$5] & -4.56 \\
4579 & 20 & 1.0$^a$     & 37.74 & 38.10 &  41.10$^d$ & 42.0$^k$ & 46.1
& -4.1 & -4.04  \\
5953 & 26 & 0.07$^a$    & $\leq$36.21 & 37.25 & -   & 44.1$^h$ & 44.9
& -0.9 & $\leq$-7.01 \\
7217 & 13 & 0.3$^a$     & 36.08 & 36.38 & 39.63$^f$ & $\sim$40$^g$ & 45.6
& [$-$5,$-$6] & -4.56  \\
\hline
\end{tabular}
\caption{Black hole masses, bolometric and Eddington luminosities for
  the seven galaxies. $^a$ assuming $M_{\rm
  bh}=(1.3\pm0.1)\times10^8$\msun$\times\left(\sigma_s/200[\kms]\right)^{4.0\pm0.3}$
  from Tremaine et al.\ 2002 with $\sigma_s$ taken from McElroy et
  al. 1995. $^b$ $\sigma_s$ taken from Ho, Filippenko \& Sargent
  (1997) and assuming
  $\sigma_s$=FWHM([O\,III])/$\sqrt{8\ln{2}}$=FWHM([N\,II])/$\sqrt{8\ln{2}}$.
  $^c$ taken from Roberts \& Warwick 2000. $^d$ taken from Ulvestad \&
  Ho 2001. $^e$ taken from Fabbiano et al.\ 1992. $^f$ taken from
  Terashima et al.\ 2002. $^g$ assuming $L_{\rm bol}$=6.7$\times
  L_{\rm x}$(2-10keV) from Ulvestad \& Ho 2001; as simplification and
  since we were interested only in a rough estimate of the bolometric
  luminosity, we assumed L$_{\rm x}$(2-10keV)$\approx L_{\rm
  x}$(0.2-4keV) allowing for half an order of magnitude
  uncertainty. $^h$ taken from Woo \& Urry (2002). $^i$ taking the
  standard Eddington equation. $^j$ assuming
  $L_{5.0}=\nu_{5.0}\times4\pi$D$^2S_\nu$ where $S_\nu$ is the
  integrated flux density from Table~\ref{radio-fluxes}. $^k$ Ho et
  al.\ 1999. $^l$ derived via D$\simeq$V$_{\rm hel}/74$\,Mpc. $^m$
  extrapolated from 15~GHz VLA observations by Nagar et al.\ (2005)
  except for NGC~2782, NGC~5953 and NGC~7217 which were taken from
  1.4~GHz VLA observations done by Becker, White \& Helfand (1995),
  Condon et al.\ (1996) or Iono, Yun \& Ho (2005). $^n$ We corrected
  the X-ray luminosities to match the distances used in this paper, if
  significantly different luminosity distances were assumed in the
  X-ray data papers. }
\label{radio-lum}
\end{table*}

\section{Nature of the radio emission: thermal or non-thermal?}
\label{sec-radio-sed} 
Based on the equation given by Condon et al.\ (1982) and taking the
deconvolved source size as an upper limit for the actual source size
and the flux density at 6~cm from Table~\ref{radio-fluxes} (except for
NGC~2782), we can estimate lower limits for the brightness temperature
in each of the seven galaxies. For most of them, the brightness
temperatures are high enough (T$_b\gtrsim 10^8\,{\rm K}$) to be
consistent with synchrotron emission given that the assumed source
sizes are only upper limits. For NGC~3147, for instance, Anderson et
al.\ (2004) derive a source size of $\lesssim$1~mas, increasing the
temperature determined from our data to $\sim$4$\times10^8$K. Given
the loose upper limit on the source size for NGC~2782, its temperature
might also be compatible with synchrotron emission. However, the lower
limit on T$_b$ virtually excludes the possibility that the radio
emission in the central $\pm$10~mas of NGC~2782 is coming from star
formation, which typically produces $\leq10^5$K, supporting the
presence of synchrotron (self-absorbed) emission from an optically
hidden AGN.

The situation for NGC~5953 and NGC~7217 might be different. Although
NGC~5953 has a very high upper limit on the source size, its
brightness temperature would only go up to $\sim$10$^7$K when taking a
source size of 1~mas, probably still too low to agree with synchrotron
self-absorption from a compact source but also too high for star
formation. However, optically thin synchrotron emission from an
(extended) jet might not disagree with the low brightness temperature;
this would also fit to the steep radio spectrum found in NGC~5953. The
continuum emission in NGC~7217 appears already slightly resolved in
the VLBI 6~cm map, leaving little leverage to increase the true
brightness temperature to $10^8\,{\rm K}$ excluding thus a
self-absorbed compact synchrotron source. In contrast to NGC5953, the
inverted radio spectra of NGC7217 are not in agreement with optically
thin synchrotron emission. Alternatively, thermal free-free emission
or electron scattered synchrotron emission may be the mechanism
responsible for the continuum similar to NGC~1068 (e.g., Roy et al.\
1998; Gallimore et al.\ 1997, 2004; Krips et al.\ 2006). Although we
see no evidence that starbursts dominate the {\it centers} of our
LLAGN, the larger VLA fluxes imply the presence of larger scale
diffuse emission that might be produced by starbursts.

\section{The fundamental plane}
\label{corr}
A highly significant relationship linking X-ray luminosity, black hole
mass, and radio luminosity has been found to hold for many black hole
systems, including both X-ray binaries and active galaxies (e.g.,
Merloni, Heinz \& Di~Matteo 2003; Falcke, K\"ording \& Markoff
2004). In Fig.~\ref{lum-mbh} \& \ref{lum-merlo}, the VLBI luminosities
at 5~GHz ($L_{5.0}$) are compared with published X-ray luminosities
(Fabbiano et al.\ 1992; Roberts \& Warwick 2000; Terashima et al.\
2002) and black hole masses (McElroy et al.\ 1995) for six of our
objects, along with literature data for $\sim$30 LINER and Seyfert
galaxies taken from Terashima et al.\ (2002), Merloni, Heinz \&
Di~Matteo (2003) and Nagar, Falcke \& Wilson (2005). In contrast to
Merloni, Heinz \& Di~Matteo (2003), and Falcke, K\"ording \& Markoff
2004, $\sim$50\% of the plotted radio fluxes were obtained with the
VLBA or VLBI, i.e., at higher angular resolution, reducing the
contamination of nuclear emission by extended components such as
(large-scale) jets or star formation.

We find that our new data are overall in good agreement with the
existing data and lie within the same empirical ``fundamental plane''
derived by Merloni, Heinz \& Di~Matteo (2003) and Falcke, K\"ording \&
Markoff (2004), even in those cases for which we used a crude estimate
for the hard X-ray luminosity. This relation has been interpreted
theoretically in terms of a disk-jet coupling, as suggested by Heinz
\& Sunyaev (2003), radiatively {\it inefficient} accretion flows, or a
combination of both. Radiatively {\it efficient} accretion flows can
be almost completely ruled out for these sub-Eddington systems
(Merloni, Heinz \& Di~Matteo 2003). The larger scatter in the
literature data compared to Heinz \& Di~Matteo (2003) and Falcke,
K\"ording \& Markoff (2004) might be caused by the uncertainty of the
assumed spectral indices that were used to extrapolate the 5~GHz
fluxes from the published 15~GHz data.

Within our sample, NGC~7217 seems to have the largest offset from the
fundamental plane. As mentioned in Section~\ref{sec-radio-sed},
NGC~7217 does not appear to be consistent with (direct) synchrotron
emission. It is unclear at this point whether its emission is produced
by electron-scattered synchrotron emission, as might be the case in
NGC~1068 (e.g., Roy et al.\ 1998; Gallimore et al.\ 1997, 2004; Krips
et al.\ 2006), or by thermal free-free radiation. The large
uncertainty of the X-ray luminosity in NGC~1068 due to the heavy
obscuration of its AGN and its large variablity (see discussion in
Gallimore et al.\ 2004) also questions the compatibility of NGC~1068
with the fundamental plane (Fig.~\ref{lum-mbh} \& \ref{lum-merlo}),
when the VLBA flux of the AGN (component S1) is plotted instead of the
lower angular resolution VLA flux adopted by Merloni, Heinz \&
Di~Matteo (2003).  The latter is certainly contaminated by emission
from the extended parts of the jet.

\section{Summary \& Discussion}
\label{sum}
This paper presents the first results of snapshot radio observations
with MERLIN and EVN of a sample of seven LLAGN taken from the NUGA
survey.  Besides compact emission, small-scale extended emission in
the form of jets is also found in these weak radio sources. The radio
fluxes decrease significantly with increasing angular resolution at
the respective wavelength, indicating the presence of extended and
diffuse emission. With one exception (NGC~5953), the radio core
components have flat to inverted spectra within the errors. A
comparison of the derived radio luminosities to the black hole mass,
estimated via the respective stellar velocity dispersions, and the
X-ray luminosities suggests correlations among these quantities even
at high angular resolution, thus supporting the existence of a
fundamental plane. However, as NGC~1068 and NGC~7217 demonstrate, this
fundamental plane could be violated if the radio emission is produced
not by direct synchrotron processes, but rather by electron scattered
synchrotron or free-free emission, even though the latter are also
expected to be closely connected to X-ray activity. Whether or not
such discrepancies reflect substantially different activity mechanisms
in NGC~1068, NGC~7217, and the remaining LLAGN in our sample has to be
further investigated. This result emphasizes the need for high angular
resolution observations of the radio emission in these sources to
verify the validity of the fundamental plane and test its limits.

\begin{acknowledgements}
Part of this work was supported by the German
\emph{Son\-der\-for\-schungs\-be\-reich, SFB\/,} project number 494.
SL acknowledges support by DGI Grant AYA 2002-03338 and Junta de
Andaluc\'ia'. We thank the referee, N.\ Nagar, for very useful and
careful comments that helped to improve the quality of the paper. We
also want to thank Dr.\ Anita Richards, Dr.\ Tom Muxlow, Dr.\ Enno
Middelberg and Dr.\ Walter Alef for their support during the
observations and data reduction. AJB acknowledges support from the
National Radio Astronomy Observatory, which is operated by Associated
Universities, Inc., under cooperative agreement with the National
Science Foundation.

\end{acknowledgements}

\end{document}